\documentclass[epj]{webofc}
\usepackage[utf8]{inputenc}
\usepackage[varg]{txfonts}   
\usepackage{booktabs}
\usepackage{xcolor}
\usepackage{slashed}
\usepackage[outdir=./pics/]{epstopdf}
\graphicspath{{./pics/}}
\definecolor{darkred}{rgb}{0.4,0.0,0.0}
\definecolor{darkgreen}{rgb}{0.0,0.4,0.0}
\definecolor{darkblue}{rgb}{0.0,0.0,0.4}
\usepackage[bookmarks,linktocpage,colorlinks,
    linkcolor = darkred,
    urlcolor  = darkblue,
    citecolor = darkgreen]{hyperref}
\usepackage{subcaption}

\graphicspath{{./pics/}}

\wocname{EPJ Web of Conferences}
\woctitle{Lattice2017}
\newcommand{\ii}{\mathrm{i}}
\DeclareMathOperator{\tr}{tr}
\DeclareMathOperator{\Tr}{Tr}

\begin{document}
%
\selectlanguage{english}
\title{%
Two-dimensional $\mathcal{N}=2$ Super-Yang-Mills Theory
}
\author{%
\firstname{Daniel} \lastname{August}\inst{1}\fnsep\thanks{Speaker, \email{daniel.august@uni-jena.de}} \and
\firstname{Bj\"orn} \lastname{Wellegehausen}\inst{1,2} \and
\firstname{Andreas}  \lastname{Wipf}\inst{1}
}
\institute{%
Theoretisch-Physikalisches Institut, Friedrich-Schiller-Universität Jena, 07743 Jena, Germany
\and
Institut für Theoretische Physik, Justus-Liebig-Universität Giessen, 35392 Giessen, Germany
}
\abstract{%
  Supersymmetry is one of the possible scenarios for physics beyond the standard model. The building blocks of this scenario are supersymmetric gauge theories. In our work we study the $\mathcal{N}=1$ Super-Yang-Mills (SYM) theory 
  	with gauge group SU(2) dimensionally 
  reduced to two-dimensional $\mathcal{N}=2$ SYM theory.
In our lattice formulation we break supersymmetry and chiral symmetry explicitly while preserving R symmetry. By fine tuning the bar-mass of the fermions in
the Lagrangian we construct a supersymmetric continuum theory. To this
aim we carefully investigate mass spectra and Ward identities, which both
show a clear signal of supersymmetry restoration in the continuum limit.
}
\maketitle
\section{Introduction}\label{sec::Introduction} 
\noindent
Supersymmetry implies a wide range of theoretical predictions and might cure fundamental drawbacks of the standard model of particle physics. Here, especially supersymmetric gauge theories play an important role in physics beyond the standard model. Since these theories are strongly coupled, non-perturbative methods are necessary to investigate the particle spectrum or expected features as confinement or chiral symmetry breaking. Beside other methods, for example functional renormalization group methods, Monte-Carlo simulations on discrete spacetime lattices play a prominent role to investigate non-perturbative aspects of (supersymmetric) gauge theories. The most simple four-dimensional supersymmetric gauge theory is $\mathcal{N}=1$ Super-Yang-Mills (SYM) theory, where extensive calculations have been done during the last years. Most of these calculations use the gauge group SU$(2)$ (see \cite{Bergner2015adz} and references therein), while calculations using SU$(3)$ are under way \cite{thiscontrib18,thiscontrib269,thiscontrib35,thiscontrib339}. At least for the gauge group SU$(2)$, it has been shown that the theory is confining, features spontaneous chiral symmetry breaking and the mass spectrum arranges in two super-multiplets as predicted by low-energy effective theories \cite{Veneziano:1982ah,Farrar:1997fn}.

Using a naive discretization of derivative operators on the lattice, supersymmetry is explicitly broken because the anticommutator of two generators $\mathcal{Q}$ of supersymmetry is proportional to the generators of translations $P_\mu$,
\begin{equation*}
	\left\{\mathcal{Q},\mathcal{Q}\right\}\sim P_\mu\,,
\end{equation*}
which do not exist on a discrete spacetime lattice. In order to recover supersymmetry in the continuum limit, it is necessary to add counterterms for every relevant operator that breaks supersymmetry explicitly in the lattice formulation. Fortunately, for $\mathcal{N}=1$ SYM in four dimensions the only counterterm that needs to be fine-tuned is the gluino condensate.
For models with extended supersymmetry, it is often possible to retain a small
subset of supersymmetry on the lattice such that a fine-tuning
may not be necessary in the continuum limit \cite{Kaplan:2003uh,Giedt:2006pd,Catterall:2009it,Joseph:2011xy,Bergner:2016sbv}. This is also true for the theory that we investigate in the present work: $\mathcal{N}=2$ SYM theory in two dimensions. This model is obtained by dimensional reduction of the aforementioned four-dimensional $\mathcal{N}=1$ SYM theory. Although a construction exists, where one nilpotent supersymmetry is realized on the lattice \cite{Sugino:2003yb,Catterall:2009it}, we use a standard discretization and identify all operators, a scalar field mass term and the gluino 
condensate, that have to be 
fine-tuned\footnote{ Deformed lattice models with an exact nilpotent
supersymmetry on the lattice could be afflicted with instabilities at 
strong couplings, see \cite{Kastner:2008zc}. }.
Our motivation to study this theory is to get further insight into fine-tuning operators for the restoration of supersymmetry in the continuum limit as well as the possibility of high precision estimates of the low energy particle spectrum which finally can
be compared to the spectrum of the four-dimensional mother-theory. In what follows we introduce the model and discuss the lattice setup in section \ref{sec::Model}, derive and show lattice Ward identities in section \ref{sec::WardIdentities} to investigate the violation of supersymmetry on the lattice and finally present our results for the mass spectrum in section \ref{sec::MassSpectrum}.

\section{$\mathcal{N}=2$ SYM theory} \label{sec::Model}
\noindent
The action of the two-dimensional $\mathcal{N}=2$ SYM-theory is given by
\begin{equation}
	\begin{aligned}
		S=\frac{1}{2 g^2}\int d^2x \tr\left\{\frac{1}{2} F_{\mu\nu}F^{\mu\nu} - \ii\,\bar{\lambda}\, \Gamma_\mu\, D^\mu\, \lambda+D_\mu\phi_m D^\mu\phi^m - \bar{\lambda}\, \Gamma_{1+m}\left[\phi^m,\lambda\right] - \frac{1}{2}\left[\phi_m,\phi_n\right]\left[\phi^m,\phi^n\right]\right\},
		\label{2dredAction}
	\end{aligned}
\end{equation}
where $F_{\mu\nu}$ is the field strength tensor, $\phi_m$ are two real scalar fields 
and $\lambda$ is a Majorana fermion, which obeys the Majorana condition
\begin{equation}
	\bar{\lambda}=\bar{\lambda}^\text{C}=\lambda^T C, \label{Majorana Condition}
\end{equation}
with charge conjugation matrix $C$. 
The fields $(A_\mu,\phi_m,\lambda)$ are all in the adjoint representation of the 
gauge group  SU$(2)$.
The four-dimensional matrices $\Gamma$ form an irreducible representation of the four-dimensional Clifford algebra. Further details on the dimensional reduction and different representations of the action can be found in a forthcoming publication \cite{August:2017}.
The action \eqref{2dredAction} is invariant under Lorentz, chiral and $R$ symmetry. The SO$(2)$ $R$ symmetry emerges naturally as a remnant of the four-dimensional Lorentz symmetry. The continuous U$(1)$ chiral symmetry is anomalously broken in the quantum theory to $\mathbb{Z}_{2N_\text{c}}$ via instantons. For a non-vanishing chiral condensate $\langle\bar{\lambda}\lambda\rangle\neq 0$, it is further broken to $\mathbb{Z}_{N_\text{c}}$.
The scalar potential $\left[\phi_m,\phi_n\right]\left[\phi^m,\phi^n\right]$ is invariant under shifts like
\begin{equation}
	\phi_1\to\phi_1 \qquad \phi_2\to\phi_2+\alpha\,\phi_1,
\end{equation}
where $\alpha$ is an arbitrary number. This leads to \emph{flat} directions in the $\left(\phi_1,\phi_2\right)$ plane, where the potential is constant. In a lattice Monte-Carlo simulation these flat directions may cause problems since 
the scalar fields may escape along these directions. To stabilize the simulations, 
the flat directions are lifted either explicitly by introducing a mass term $m_\text{S}^2\,\phi^2$ for the scalars and taking the limit $m_\text{S}\to 0$ at the end, 
or dynamically by quantum corrections. We have demonstrated earlier that 
the latter takes place \cite{August:2016orf} and no explicit mass term for the
scalars is needed in simulations.

The lattice formulation of this theory is based on \cite{Nicolai:1978vc,Montvay:2001ry}. We use Wilson fermions and therefore break the chiral symmetry explicitly. In contrast to four dimensions, the mass term $m_\text{F}\,\bar{\lambda}\lambda$ is irrelevant in two dimensions, making a fine-tuning in the UV unnecessary to reach the correct continuum limit in the IR. However, at finite lattice spacing an effective fermion 
mass is generated by discretization errors and should be canceled by an explicit fermion mass term. Another motivation to introduce a fermion mass term arises from the four-dimensional SYM theory, where supersymmetry is automatically restored if chiral symmetry is realized in the continuum limit \cite{Curci:1986sm}. For both reasons, we fine-tune the bare fermion mass to its critical value defined by a vanishing renormalized gluino mass.

In our simulations we use the tree-level improved L\"uscher-Weisz gauge action \cite{Luscher:1984xn}. The scalar fields will be discretized as on-site variables 
and the covariant lattice-derivative is given by
\begin{equation}
	D_\mu \phi_x=\phi_{x+e_\mu} U_{x,\mu}^\text{Adj}-\phi_x\,,
\end{equation}
where the link variables $U^\text{Adj}$ are in the adjoint representation of the gauge group.
Analytic calculations in \cite{Sugino:2003yb,Suzuki:2005dx} show that we have to add a scalar mass term $m_\text{S}^2\,\phi^2$ to cancel contributions in the one-loop effective lattice action that are not present in the continuum formulation. For practical reasons we decided not 
to fine-tune the scalar mass but to fix it to the analytically known result $m^2_\text{S}=0.65948255(8)$ \cite{Suzuki:2005dx}.

In order to check for finite volume effects, we performed simulations on different lattice sizes $32\times16,48\times 24,64\times 32$ and $96\times 48$. We observe that the 
Ward identities are rather insensitive to the size of the lattice. On the other hand, 
the meson ground state masses are much affected by finite size effects and make an extrapolation to  infinite volume necessary. Most of our simulations are done on the $64\times 32$ lattice which is a good compromise between finite size effects and simulation speed. On this lattice we generated $1000$ to $20\,000$ statistically independent configurations for different values of the inverse gauge coupling $\beta$. The setup for these lattices is summarized in Table \ref{tab::latticespacing}.

In order to determine the lattice spacing, we extract the static quark-antiquark potential in the fundamental representation from Wilson loops and extrapolate it to the chiral limit. To compare our results to usual QCD lattice data, we employ the Sommer scale \cite{Sommer:1993ce} and define a lattice spacing in \emph{physical} units. The results for three different values of the inverse gauge coupling
\begin{equation}
	\beta=\frac{1}{a^2 g^2}
\end{equation}
are also shown in Table \ref{tab::latticespacing}. With increasing inverse gauge coupling $\beta$, the lattice spacing becomes smaller and the (in two dimensions dimensionful) gauge coupling $g$ stays constant. This shows that the continuum limit is reached
for $\beta \to \infty$.
\begin{table}[htb]
  \small
  \centering
  \begin{tabular}{ccccc}\toprule
  $T\times L$ & $\beta$ & $a\,[\text{fm}]$ & $g^{-2}=\beta\, a^2\,[\text{fm}^2]$ & \# of configurations  \\\midrule
  $64\times 32$ & 17.0 & 0.0591(16) & 0.0595(31) & $20\,000$\\
  $64\times 32$ & 15.5 & 0.0618(18) & 0.0592(35) & $20\,000$\\
  $64\times 32$ & 14.0 & 0.0650(25) & 0.0592(46) & $20\,000$\\\bottomrule
  \end{tabular}
  \caption{Lattice setup and results for the lattice spacing $a$ and the gauge coupling $g$ for different values of $\beta$, after extrapolating to the chiral limit. \label{tab::latticespacing}}
\end{table}
\section{Ward Identities} \label{sec::WardIdentities}
\noindent
To check for restoration of supersymmetry in the continuum and infinite volume limit we
investigate several Ward identities. Ward identities are relations between different 
correlations functions, which follow from supersymmetry. The derivation makes use of the fact that the super charge $\mathcal{Q}$ annihilates the vacuum, leading to the identity
\begin{equation}
	\langle \mathcal{Q}\,\mathcal{O} \rangle=0,
\end{equation}
where $\mathcal{O}$ is an arbitrary operator. Here, we use
\begin{equation}
	\mathcal{O}(x)=\Tr\left[\bar{\lambda}_b(x)\left(\Gamma_{\mu\nu}\right)^b_a F^{\mu\nu}(x)\right]
\end{equation}
that leads to the bosonic Ward identity
\begin{equation}
	\frac{\beta}{V} \,\langle S_\text{B} \rangle=-\frac{3}{16}\big\langle\,\ii\bar{\lambda}(x)\,\slashed{D}\,\lambda(x)
	\big\rangle=\frac{9}{2},
\end{equation}
where $S_\text{B}$ is the bosonic part of the action. This identity is a continuum identity, which makes use of unbroken supersymmetry. On the lattice supersymmetry is broken and we have to modify the Ward identity according to
\begin{equation}
	\langle \mathcal{Q}\,\mathcal{O} \rangle+\langle \mathcal{Q}\,\chi_\text{f} \rangle+\langle \mathcal{Q}\,\chi_\text{s} \rangle+\langle \mathcal{Q}\,\chi_\text{L} \rangle=0.
\end{equation}
Here $\chi_\text{f}$ is the contribution that arises from the fermion mass term, $\chi_\text{s}$ the contribution from the scalar mass term and $\chi_\text{L}$ is the effect of the lattice regularisation. As expected, a dimensional analysis reveals, that at tree-level the term $\langle \mathcal{Q}\,\chi_\text{L} \rangle$ does not contribute in the continuum limit. For higher loop orders, we use the results in \cite{Suzuki:2005dx}, where the authors demonstrated that the only UV divergent diagram corresponds to the one-loop scalar two-point function. Hence only this term may suffer from lattice artefacts in the continuum limit. Applying this result to the Ward identity, we have to add the scalar mass term to the bosonic action, arriving at
\begin{equation}
	W=\beta \left(V^{-1}\langle S_\text{B} \rangle+\langle m^2_\text{S} \,\phi^2 \rangle+\langle m_\text{F}\,\mathcal{Q}\,\bar{\lambda}\lambda \rangle+\langle m^2_\text{S}\,\mathcal{Q}\,\phi^2 \rangle\right)= \frac{9}{2}.\label{WardCor}
\end{equation}
To investigate the effect of the corrections we split these four contributions into two parts, the continuum bosonic action with the scalar mass, denoted by $W_\text{B}$, and the remainder denoted by $W_\text{corr}$,
\begin{equation}
 W_\text{B}=\left\langle\beta\, \left(V^{-1}\,S_\text{B} +m^2_\text{S}\, \phi^2\right)\right\rangle\,,\quad W_\text{corr}=\left\langle\beta\,\left(m_\text{F}\,\mathcal{Q}\,\bar{\lambda}\lambda+m^2_\text{S}\,\mathcal{Q}\,\phi^2\right)\right\rangle. 
\end{equation}

\begin{figure}[htb]
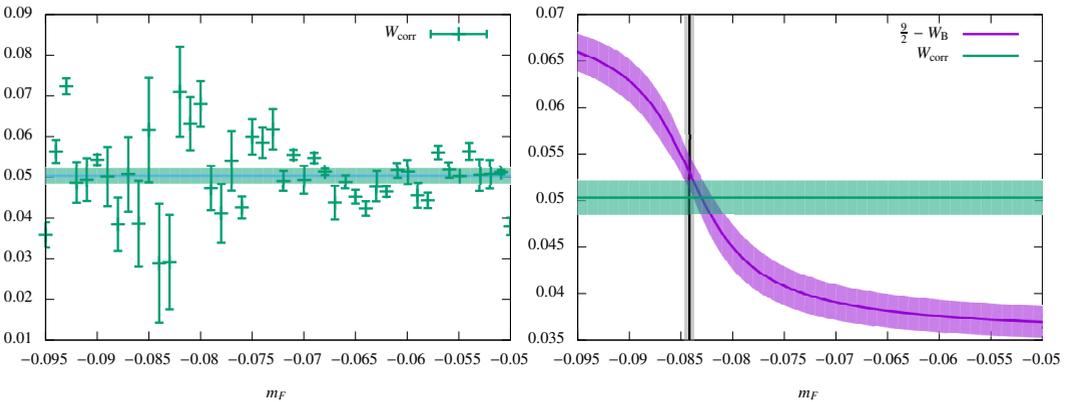
 
  		\scalebox{0.87}{\input{pics/FullCorrections185.tex}}
  		\scalebox{0.87}{\input{pics/BosonicActionCorrections185.tex}}
  \caption{Left panel: Lattice corrections to the bosonic Ward identity as function of the bare fermion mass $m_\text{F}$ at $\beta=18.5$ together with its average. Right panel: The averaged lattice correction is compared to the uncorrected bosonic Ward identity at $\beta=18.5$. The vertical line denotes the critical fermion mass and the shaded regions indicate the 68\% confidence interval for the critical value.}
  \label{fig::BosonicWard}
\end{figure}
In the left panel of Figure \ref{fig::BosonicWard}, we show the lattice corrections plotted as a function of the bare fermion mass. Within statistical errors we did not observe any dependence on $m_\text{F}$ and therefore we take the average value over the $m_\text{F}$-range as correction to the Ward identity. In the right panel of Figure \ref{fig::BosonicWard} we show expectation value $9/2-W_\text{B}$ (fitted to an $\arctan$ function) together with the corrections and the critical fermion mass at which chiral symmetry is restored. Close to the chiral limit both curves intersect and the Ward identity \eqref{WardCor} is fulfilled while it is violated away from the critical point. Since all relevant lattice
operators are taken into account, the observed violation of the Ward identity
is due to irrelevant operators, present at finite lattice size and spacing. These irrelevant operators depend on
the renormalized fermion mass, which is zero at the critical fermion mass. Hence we fulfill the Ward identity in the chiral limit. This means that fine-tuning of the 
fermion mass will reduce lattice artefacts and push the results closer to the supersymmetric limit. 
This agrees with similar observations made for four-dimensional $\mathcal{N}=1$ SYM theory 
\cite{Curci:1986sm}. Note that this observation is consistent with the works 
\cite{Sugino:2003yb,Suzuki:2005dx}, because in the continuum limit, the critical fermions 
mass vanishes.

\section{Mass Spectrum}\label{sec::MassSpectrum}
\noindent
The effective field theory for the four-dimensional $\mathcal{N}=1$ theory predicts two particle multiplets, the Veneziano-Yankielowicz (YM) multiplet \cite{Veneziano:1982ah} and the Farrar-Gabadadze-Schwetz (FGS) multiplet \cite{Farrar:1997fn}. Dimensional reduction leads to particle multiplets of the two-dimensional theory, shown in Table \ref{tab::MassSpectrum}.
\begin{table}[htb]
\begin{subtable}[b]{0.05\textwidth}
~
\end{subtable}
\begin{subtable}[b]{0.35\textwidth}
  \small
  \begin{tabular}{lll}\toprule
  particle & spin & name  \\\midrule
  $\bar{\lambda}\,\Gamma_5\,\lambda$ & 0 & $a-\eta$ \\
  $\bar{\lambda}\lambda$ & 0 & $a-f$ \\
  $F_{\mu\nu}\,\Sigma^{\mu\nu}$ & $\frac{1}{2}$ & gluino-glueball \\\bottomrule
  \end{tabular}
\end{subtable}
\begin{subtable}[b]{0.08\textwidth}
~
\end{subtable}
\begin{subtable}[b]{0.35\textwidth}
  \small
  \begin{tabular}{lll}\toprule
  particle & spin & name  \\\midrule
  $\left[\phi_1,\phi_2\right]F_{\mu\nu}$ & 0 & glue-scalarball \\
  $F_{\mu\nu}F^{\mu\nu},\left[\phi_1,\phi_2\right]^2$ & 0 & $0^+$-glueball, scalarball \\
  $F_{\mu\nu}\,\Sigma^{\mu\nu}$ & $\frac{1}{2}$ & gluino-glueball \\\bottomrule
  \end{tabular}
\end{subtable}
\caption{Multiplets obtained from dimensional reduction of the VY multiplet(left) and the FGS multiplet (right). The first multiplet contains a scalar meson, a pseudoscalar meson and a gluino-glueball, while the second multiplet consists of scalar-, glue- and glue-scalarballs and a gluino-glueball.\label{tab::MassSpectrum}}
\end{table}
As in four dimensions, the first multiplet consists of a scalar meson $f$, a pseudoscalar meson $\eta$ and a spin $1/2$ gluino-glueball. The second multiplet contains glueballs, bound states of scalar fields (scalarballs) and bound states between gluons and scalar fields (glue-scalarballs). Additionally it contains also a gluino-glueball. 
The masses of these particles in the continuum limit and with restored supersymmetry is the main focus of our work.

We also monitor the adjoint pion mass to control the chiral (and supersymmetric) limit 
of the theory. Since we have only a single fermion flavour
(in the four-dimensional theory), we define the pion correlation 
function as the connected part of the adjoint $\eta$-meson correlation function (partially 
quenched approximation). This is motivated by the four-dimensional theory 
\cite{Veneziano:1982ah}. The reason to introduce this quantity is, that in four 
dimensions, the square of the pion mass is proportional to the gluino mass 
\cite{Donini:1997gf}. This allows to define the chiral limit as $m_\pi\to 0$ with a 
reliable extrapolation to zero gluino mass. Of course we have to check that this method is 
applicable in two dimension.
\begin{figure}[htb]
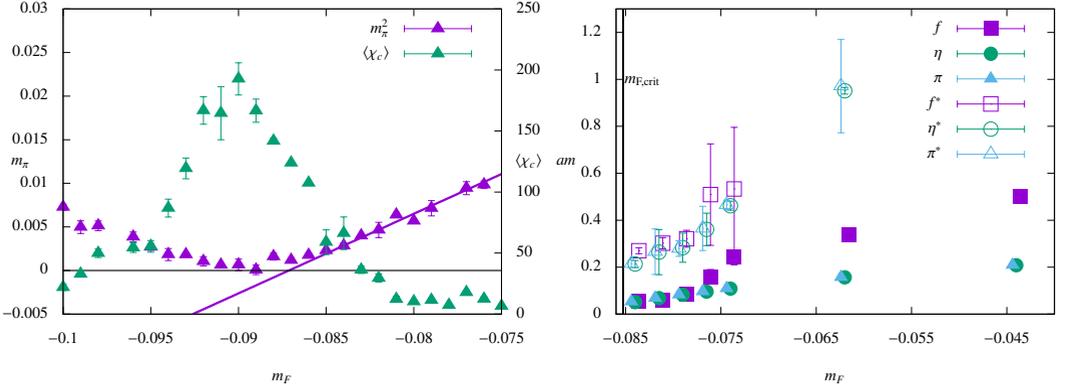

  		\scalebox{0.85}{\input{pics/pionmasssusc2.tex}}\hskip5mm
  		\scalebox{0.85}{\input{pics/Massvglbeta17.tex}}
  \caption{Left panel: Pion mass and chiral susceptibility as a function of the fermion mass $m_\text{F}$. The critical fermion mass obtained from the fit to the gluino mass is $m_\text{F,crit}= -0.0853(1)$ while we get $m_\text{F,crit}= -0.0896(22)$ from the peak of the chiral susceptibility. Right panel: Ground state masses for the $\eta$- and $f$-meson and for the pion for several bare fermion masses $m_\text{F}$ at $\beta=17.0$.}
  \label{fig::CriticalMass}
\end{figure}
In Figure \ref{fig::CriticalMass} we compare the critical fermion mass (left panel) to the peak of the chiral susceptibility, which we expect to coincide in the continuum and supersymmetric limit. We find for the former $m_\text{F,crit}= -0.0853(1)$  and for the latter $m_\text{F,crit}= -0.0896(22)$, showing that both definitions agree reasonable well. We observe also that the square of the pion mass is a linear function in the bare fermion mass in the region $\left[-0.085,-0.06\right]$, which is expected in four dimensions but also holds in two dimensions. Subsequently we use the pion mass to determine the critical fermion mass at which chiral symmetry gets restored in the continuum limit.

The ground state masses of the $\eta$- and $f$-meson and the pion are plotted in Figure \ref{fig::CriticalMass} (right panel) as function of the bare fermion mass. We observe, that the masses for the $\eta$-meson and the pion are identical. The reason is that the disconnected part contributes
at most 10 percent to the full $\eta$-meson correlation function. Hence the difference of
both correlation functions lies within statistical errors and consequently their masses 
are degenerate. This result holds in the chiral and infinite volume limit for
various values of the inverse gauge coupling. We conclude that the $\eta$-meson is 
massless in the continuum limit and we are left with a theory without mass gap, in 
agreement with analytical calculations \cite{Witten:1995im,Fukaya:2006mg}. The $f$-meson 
is significantly heavier away from the chiral limit, which is expected because chiral 
symmetry is broken explicitly. However, close to the critical fermion mass it 
degenerates with the $\eta$ mass, indicating restoration of chiral symmetry. For the masses of the 
excited mesons, we find agreement in the chiral limit too. A simulation directly at the 
critical fermion mass shows that the $f$, $\eta$ and pion correlation functions are indeed 
identical within statistical uncertainties \cite{August:2016orf}. Combining these results, 
we conclude that the $\eta$- and $f$-meson lie indeed in the same multiplet as expected 
and become massless in the continuum limit.

\begin{figure}[htb]
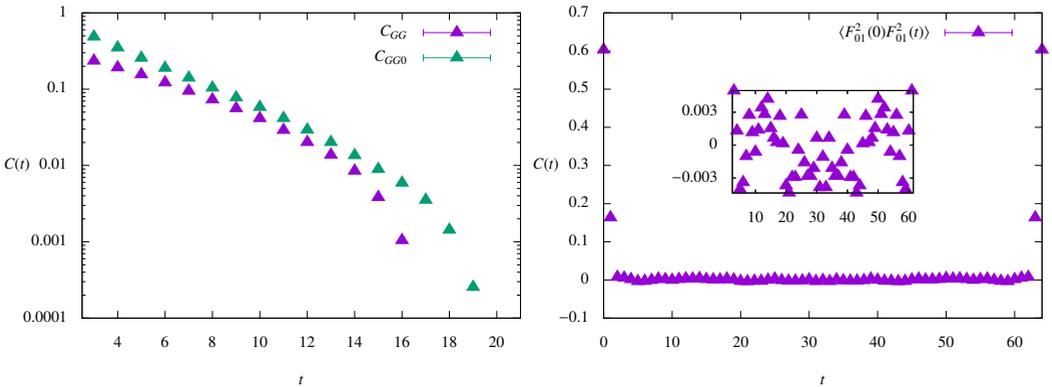
 
  		\scalebox{0.85}{\input{pics/GluinoGlueCorr.tex}}
  		\scalebox{0.85}{\input{pics/GlueBallCorr.tex}}
  \caption{Left: Correlation functions for the gluino-glueball contracted with the identity matrix ($C_\text{GG}$) and with $\Gamma_0$ ($C_\text{GG0}$). Right: Glueball correlation function with zoom into the region $t\in \left[3,61\right]$.}
  \label{fig::GluinoGlue}
\end{figure}

In Figure \ref{fig::GluinoGlue} (left panel) we show two correlation functions associated with the gluino-glueball. 
Since the corresponding interpolating operators transforms like a spin $1/2$ field, we can insert the four dimensional gamma matrices in the correlation function to project onto different contributions. The only two independent possibilities are
\begin{equation}
	C_\text{GG}=\langle\bar{\mathcal{O}}(x)\,\mathcal{O}(y)\rangle,\qquad C_\text{GG0}=\langle\bar{\mathcal{O}}(x)\,\Gamma_0\,\mathcal{O}(y)\rangle, \qquad \text{with} \qquad \mathcal{O}(x)=F_{\mu\nu}(x)\,\Gamma^{\mu\nu}\,\lambda(x).
\end{equation}
For the calculation of the correlation functions we employ 300 steps of STOUT smearing for
the link variables and five steps of Jacobi smearing for the fermion sources. The masses, 
obtained by a $\cosh$- respectively $\sinh$-fit are $0.231(8)$ and $0.297(7)$. They are 
almost independent of the bare fermion mass and are comparable to the excited state 
masses of the mesons in the chiral limit. Unfortunately at present we are 
unable to decide whether these values belong to the ground state or first excited state 
of the gluino-glueball.

Finally we investigate the glueball, scalarball and glue-scalarball correlation functions. The correlation function for the glueball does not show any correlation for $t>2$, which can be seen in Figure \ref{fig::GluinoGlue} (right panel). 
This is reminiscent of the two dimensional pure Yang-Mills theory, where Wilson loops interact only via overlapping or touching plaquettes. Since we use the clover plaquette in the interpolating operator for the glueball, we would expect interaction for $t<3$ while for $t\geq3$ the correlation function for the glueball should vanish. The same behaviour is also observed for the scalarball and the glue-scalarball. We conclude, that all three particles are in the same multiplet, the mass of which tends to infinity and therefore decouples from the theory.

\section{Conclusion}

We have employed the conventional Wilson fermion lattice discretization to simulate 
two-dimensional $\mathcal{N}=2$ SYM theory, which can be derived by dimensional
reduction of four-dimensional $\mathcal{N}=1$ SYM theory. The reduction maintains the chiral properties of the theory. We identified two parameters, 
the fermion and the scalar 
mass, that have to be fine-tuned to define a chiral and supersymmetric continuum limit. In 
order to monitor the violation of supersymmetry due to its explicit breaking at finite 
lattice spacing we derived lattice Ward identities. With simulations at different gauge 
couplings and on different lattice sizes we confirmed that supersymmetry and chiral 
symmetry are indeed restored in the continuum and infinite volume limit. Furthermore, it 
turns out that the renormalized gluino mass is the source of explicit supersymmetry 
breaking terms on the lattice, revealing a strong connection between chiral and 
supersymmetry. This is also expected from simulations in four dimensions.
Contrary to the four-dimensional model \cite{Bergner2015adz}, the mesons are massless 
in the continuum limit. This is in accordance with analytical calculations 
\cite{Witten:1995im,Fukaya:2006mg} and previous numerical studies 
\cite{Antonuccio:1998mq,Harada:2004ck}. The glueballs are infinitely heavy and decouple 
from the theory, similarly as observed in two-dimensional pure Yang-Mills theory. 
The glueballs are expected 
to be in the same multiplet with scalarballs and glue-scalarballs and we have shown that 
these states, due to the supersymmetry, decouple too. The gluino-glueball state which is 
present in both multiplets, did not seem to fit into the meson multiplet. Concerning
the glueball multiplet we cannot make a definite statement,
because the gluino-glueball in this multiplett should decouple and thus
is not seen on the lattice.

\section*{Acknowledgments}
B.~W. and D.~A. have been supported by the DFG Research 
Training Group 1523/2 “Quantum and Gravitational Fields" and in part by the DFG (Grant Wi 777/11).
B.~W. thanks the Helmholtz International Center for FAIR within the LOEWE initiative of the State of Hesse
for support.

\bibliography{lattice2017}

\end{document}